\documentclass[aps,prd,preprint,superscriptaddress,showpacs,floatfix,nofootinbib]{revtex4}

\usepackage{graphicx}
\usepackage{amsmath,amssymb}
\usepackage{amsfonts}
\usepackage{verbatim}
\usepackage{latexsym}
\usepackage{amsthm}
\usepackage{amsbsy}
\usepackage{multirow}

\newcommand{\bmat}{\left(\begin{array}}
\newcommand{\emat}{\end{array}\right)}
\newcommand{\beq}{\begin{equation}}
\newcommand{\eeq}{\end{equation}}

\def\yzero{\smash{\hbox{$y\kern-4pt\raise1pt\hbox{${}^\circ$}$}}}

\def\-{\hphantom{-}}

\def\s2{\frac{1}{\sqrt2}}

\def\be{\begin{equation}}
\def\ee{\end{equation}}
\def\bea{\begin{eqnarray}}
\def\eea{\end{eqnarray}}

\def\IF{\relax{\rm I\kern-.18em F}}
\def\II{\relax{\rm I\kern-.18em I}}
\def\IP{\relax{\rm I\kern-.18em P}}

\def\Dsl{\,\raise.15ex\hbox{/}\mkern-13.5mu D} 

\def\IC{\bf C}
\def\IZ{\bf Z}

\def\z2z2{$\IC^3/(\IZ_2\times\IZ_2)$}

\def\s{\sigma}

\def\z{\zeta}

\def\bo{{\raise-.3ex\hbox{\large$\Box$}}}               
\def\face{{\raise.2ex\hbox{$\displaystyle \bigodot$}\mskip-2.2mu \llap {$\ddot
        \smile$}}}                                      

\def\leftrightarrowfill{$\mathsurround=0pt \mathord\leftarrow \mkern-6mu
        \cleaders\hbox{$\mkern-2mu \mathord- \mkern-2mu$}\hfill
        \mkern-6mu \mathord\rightarrow$}       
\def\dvec#1{\vbox{\ialign{##\crcr
        \leftrightarrowfill\crcr\noalign{\kern-1pt\nointerlineskip}
        $\hfil\displaystyle{#1}\hfil$\crcr}}}           

\def\beq{\begin{equation}}
\def\eeq{\end{equation}}

\def\beqx{\begin{displaymath}}
\def\eeqx{\end{displaymath}}

\def\beqa{\begin{eqnarray}}
\def\eeqa{\end{eqnarray}}

\begin{document}

\title{ Three-Dimensional ${\cal N}\geq 5$ Superconformal Chern-Simons Gauge
Theories And Their Relations }

\author{Tianjun Li}

\affiliation{Institute of Theoretical Physics, 
Chinese Academy of Sciences, Beijing 100080, P. R. China }

\affiliation{George P. and Cynthia W. Mitchell Institute for
Fundamental Physics, Texas A$\&$M University, College Station, TX
77843, USA }

\date{\today}

\begin{abstract}

We propose three-dimensional ${\cal N}=6$ superconformal $U(N)\times U(M)$ 
and $SU(N)\times SU(N)$ Chern-Simons gauge theories with two pairs of 
bifundamental chiral superfields in 
the $(\mathbf{N}, \mathbf{M})$ and $(\mathbf{\overline{N}}, \mathbf{\overline{M}})$
representations and in the $(\mathbf{N}, \mathbf{N})$ and  
$(\mathbf{\overline{N}}, \mathbf{\overline{N}})$
representations, respectively.
We also propose the superconformal $U(1)\times U(1)$ gauge 
theories that have $n$ pairs of bifundamental chiral superfields with 
 $U(1)\times U(1)$ charges $({\mathbf{\pm1}}, {\mathbf{\mp1}})$ 
 or  $({\mathbf{\pm1}}, {\mathbf{\pm1}})$. Although these $U(1)\times U(1)$ 
gauge theories have global symmetry  $SU(2n)$, 
the R-symmetry is  $SO(6)$  for $n=2$, and might be $SO(2n)$ or $SO(2n+1)$ 
for $3 \leq  n \leq 8$.
In addition,  we show that from either the generalized ABJM theories, or 
our $U(N)\times U(M)$ theories, or the ${\cal N}=5$ superconformal 
$O(N) \times USp(2M)$ gauge theories, we can derive
all the ${\cal N} \geq 5$ superconformal Chern-Simons gauge theories
except the ${\cal N}=5$ superconformal $G_2\times USp(2)$  gauge theory 
and our $U(1)\times U(1)$  gauge theories with $n\not= 2$ and 4.
Furthermore, we derive the three-dimensional  ${\cal N}=8$ 
superconformal $U(1) \times U(1)$ gauge theory from the BLG theory,
and study the corresponding moduli space. With the novel
Higgs mechanism in the unitary gauge, we suggest that it
 describes a D2-brane and a decoupled D2-brane.

\end{abstract}


\preprint{MIFP-08-24}

\maketitle



\section{Introduction}

A Lagrangian description for the worldvolume of multiple M2-branes
is very important to understand the M-theory. Especially, the 
superconformal field theory  on the worldvolume of multiple 
M2-branes is dual to the M-theory on $AdS_4\times S^7$ in the AdS/CFT
correspondence. The superconformal Chern-Simons gauge theories without
Yang-Mills kinetic terms have been studied for this 
purpose~\cite{Schwarz:2004yj}, but, they did not have enough 
supersymmetry. Along this approach, 
Barger and Lambert~\cite{Bagger:2006sk, Bagger:2007jr}, as well as 
Gustavasson~\cite{Gustavsson:2007vu} (BLG) have successfully 
constructed three-dimensional ${\mathcal N}=8$ superconformal 
Chern-Simons gauge theory 
with manifest $SO(8)$ R-symmetry based on three algebra.
And then there is intensive research on three-dimensional
superconformal gauge theories and their relations to the 
low energy theory on M2-branes~\cite{Mukhi:2008ux, VanRaamsdonk:2008ft,
Lambert:2008et, Distler:2008mk, Ho:2008bn, Papadopoulos:2008sk, 
Gauntlett:2008uf, Li:2008ya}. Although the 
BLG theory is expected to describe any number of M2-branes, 
its gauge group can only be $SO(4)$ for the positive definite 
metric~\cite{Ho:2008bn, Papadopoulos:2008sk, Gauntlett:2008uf}. 
At the Chern-Simons level one, the BLG 
SO(4) gauge theory describes two M2-branes on a $R^8/Z_2$ 
orbifold~\cite{Lambert:2008et, Distler:2008mk}.

To generalize the BLG theory so that
it can describe an arbitrary number of M2-branes,
Aharony, Bergman, Jafferis and Maldacena (ABJM)
have constructed three-dimensional ${\cal N}=6$ superconformal 
Chern-Simons gauge theories with groups $U(N)\times U(N)$ and 
$SU(N)\times SU(N)$~\cite{Aharony:2008ug} (For Chern-Simons gauge 
theories with ${\cal N}=3$ and $4$ supersymmetries, see 
Refs.~\cite{Gaiotto:2007qi, Gaiotto:2008sd}). Using
brane constructions they argued that the  $U(N)\times U(N)$
theory at Chern-Simons level $k$ describes the low-energy 
limit of $N$ M2-branes on a $C^4/Z_k$ orbifold. In particular,
 for $k=1$ and $2$, ABJM conjectured that their theory
describes  the $N$ M2-branes respectively in the flat space and 
on a $R^8/Z_2$ orbifold, and then might have ${\cal N}=8$ supersymmetry.
Especially, the $SU(2)\times SU(2)$ theory has enhanced 
$SO(8)$ R-symmetry and 
is the same as the BLG theory~\cite{Aharony:2008ug}. And it was proposed
that the $U(2)\times U(2)$  theory might also have the enhanced ${\cal N}=8$ 
supersymmetry at Chern-Simons level one and two~\cite{Klebanov:2008vq}.
Moreover, the ABJM theories can be easily generalized to have the 
$U(N)\times U(M)$ gauge symmetry~\cite{Aharony:2008ug, Hosomichi:2008jb}.
Thus, in this paper, we will define the ABJM theories as the generalized
ABJM theories that can have gauge groups $U(N)\times U(M)$ or 
$SU(N)\times SU(N)$. In addition, the superconformal Chern-Simons 
gauge theories with ${\cal N}=5$ and 6 supersymmetries have been 
classified recently~\cite{Hosomichi:2008jb, Bagger:2008se, 
Schnabl:2008wj, Bergshoeff:2008bh, Aharony:2008gk}:
the gauge groups for ${\cal N}=6$ supersymmetry are 
$SU(N)\times SU(N)$, $U(N)\times U(M)$,
 and $U(1)\times USp(2N)$; and the gauge groups for ${\cal N}=5$ 
supersymmetry are $O(N) \times USp(2M)$ and $G_2\times USp(2)$.
We would like to emphasize that for all the ${\cal N}=6$ superconformal  
 $U(N)\times U(M)$ and $SU(N)\times SU(N)$ 
Chern-Simons gauge theories that have been
 constructed so far~\cite{Aharony:2008ug, Hosomichi:2008jb, Schnabl:2008wj}, 
there are two pairs of bifundamental chiral superfields in 
the $(\mathbf{N}, \mathbf{\overline{M}})$ and  
$(\mathbf{\overline{N}}, \mathbf{M})$ representations, and in
the $(\mathbf{N}, \mathbf{\overline{N}})$ and  
$(\mathbf{\overline{N}}, \mathbf{N})$ representations,
respectively.

As we know, the gauge symmetry for $N$ stacks of D-branes in Type II
theories is $U(N)$. With orientifold actions, we can obtain the 
$SO(M)$ or $USp(2N)$ gauge symmetry for the D-branes on the 
top of orientifold planes via orientifold projections. Moreover,
from  the $SO(M)$ or $USp(2N)$ gauge symmetry for the D-branes on 
the top of orientifold planes, we can obtain back
the $U(N)$ gauge symmetry by putting the same number of D-branes 
on all the orientifold images (for an example, see Ref.~\cite{Cvetic:2004nk}).
Therefore, it seems to us that the generic three-dimensional  
${\cal N} \geq 5$ superconformal Chern-Simons gauge theories may be derived 
from the ${\cal N}=6$ superconformal $U(N)\times U(M)$ gauge theories
and the ${\cal N}=5$ superconformal $O(N)\times USp(2M)$ gauge theories.

In this paper, we first propose the three-dimensional 
${\cal N}=6$ superconformal $U(N)\times U(M)$ gauge theories
with two pairs of bifundamental chiral superfields in 
the $(\mathbf{N}, \mathbf{M})$ and  
$(\mathbf{\overline{N}}, \mathbf{\overline{M}})$ representations,
and the ${\cal N}=6$ superconformal $SU(N)\times SU(N)$
gauge theories with two pairs of bifundamental chiral superfields in 
the $(\mathbf{N}, \mathbf{N})$ and  
$(\mathbf{\overline{N}}, \mathbf{\overline{N}})$ representations.
For our $SU(N)\times SU(N)$ theory, we obtain the BLG theory when
$N=2$. Moreover, we  propose the superconformal $U(1)\times U(1)$ gauge 
theories that have $n$ pairs of bifundamental chiral superfields with 
the $U(1)\times U(1)$ charges $({\mathbf{+1}}, {\mathbf{-1}})$ and 
$({\mathbf{-1}}, {\mathbf{+1}})$, or the $U(1)\times U(1)$ charges 
$({\mathbf{+1}}, {\mathbf{+1}})$ 
and $({\mathbf{-1}}, {\mathbf{-1}})$. These theories have global 
symmetry  $SU(2n)$, and it seems to us that the R-symmetry is 
$SO(6)$ for $n=2$, and might be
 $SO(2n)$ or $SO(2n+1)$ for $3 \leq n \leq 8$.

With the similar mechanism for Wilson line gauge symmetry breaking,
 we show that in the ABJM $U(N)\times U(M)$ theories and 
our $U(N)\times U(M)$ theories, the ${\cal N}=6$ superconformal 
$U(N')\times U(M')$ 
Chern-Simons gauge theories can be obtained from the ${\cal N}=6$ superconformal 
$U(N)\times U(N)$ Chern-Simons gauge theories,  and the ${\cal N}=6$ 
superconformal $U(N')\times U(N')$ Chern-Simons gauge theories can be obtained 
from the ${\cal N}=6$ superconformal $U(N)\times U(M)$ 
Chern-Simons gauge theories, where $N' \leq N$,  $M' \leq N$, and $N' \leq M$.
In addition, we prove that the ${\cal N}=5$ superconformal 
$O(N) \times USp(2M)$ Chern-Simons gauge theories can be derived from 
the  ABJM $U(N)\times U(2M)$ theories and our $U(N)\times U(2M)$ theories.
Also, both the ABJM $U(N)\times U(M)$ theories and 
our $U(N)\times U(M)$ theories can be derived from the 
${\cal N}=5$ superconformal $O(2N) \times USp(2M)$ Chern-Simons gauge 
theories. Moreover, we point out that the  ${\cal N}=5$ superconformal
$O(2)\times USp(2N)$ gauge theories have enhanced $SO(6)$ R-symmetry,
and become the ${\cal N}=6$ superconformal $U(1) \times USp(2N)$
gauge theories. And it seems to us that the ${\cal N}=5$ superconformal 
$G_2\times USp(2)$ Chern-Simons gauge theory might be obtained from 
the  ${\cal N}=5$ superconformal $O(7) \times USp(2)$ Chern-Simons gauge
theory since $G_2$ is a special maximal subgroup of $SO(7)$.

Furthermore, we derive the three-dimensional  ${\cal N}=8$ superconformal 
$U(1) \times U(1)$ gauge theory from the BLG theory, which can be
considered as our above superconformal $U(1)\times U(1)$ gauge theory
with four pairs of chiral superfields whose 
 $U(1)\times U(1)$ charges are $({\mathbf{+1}}, {\mathbf{-1}})$ and 
$({\mathbf{-1}}, {\mathbf{+1}})$. And the moduli space has been studied 
in details. With the novel Higgs mechanism in the unitary gauge, we show 
explicitly that this theory may describe a D2-brane and a decoupled D2-brane,
and we present the concrete physics picture as well.
Although the R-symmetry in our theory 
should be $SO(8)$, the global symmetry is indeed $SU(8)$. 
We emphasize that this
superconformal $U(1) \times U(1)$  gauge theory 
is different from the ABJM $U(1) \times U(1)$ theory since we have 
eight bifundamental chiral superfields.

This paper is organized as follows. In Section II, we propose
the three-dimensional ${\cal N}=6$ superconformal $U(N)\times U(M)$ 
and $SU(N)\times SU(N)$ gauge theories, and the ${\cal N} \geq 6$
superconformal $U(1)\times U(1)$ gauge theories. 
In Section III, we study the relations among
the ${\cal N} \geq 5$ superconformal Chern-Simons gauge theories.
In Section IV, we consider three-dimensional  
${\cal N}=8$ superconformal $U(1) \times U(1)$ gauge theory in details.
Our discussion and conclusions are given in Section V.

\section{New Three-Dimensional Superconformal Chern-Simons Gauge Theories}

In this Section, we will propose the three-dimensional 
${\cal N}=6$ superconformal $U(N)\times U(M)$ and 
$SU(N)\times SU(N)$ gauge theories, and the ${\cal N} \geq 6$
superconformal $U(1)\times U(1)$ gauge theories. The detail study
of our theories will be presented elsewhere.
Although the Chern-Simons level $k$ for the ${\cal N}=6$ superconformal 
$U(N)_k\times U(M)_{-k}$ gauge theories should be equal to or larger 
than $|N-M|$~\cite{Aharony:2008gk}, {\it i.e.}, $k \geq |N-M|$, 
we will neglect this subtlety for simplicity in the following discusssions.

\subsection{ ${\cal N}=6$ Superconformal $U(N) \times U(M)$ Theories}

Following the ABJM construction~\cite{Aharony:2008ug}, we consider 
the ${\cal N}=3$ supersymmetric $U(N)\times U(M)$ theories. Starting from 
the field content of an ${\cal N}=4$ supersymmetric theory, we need to 
add to the $U(N)$ and $U(M)$ vector multiplets the auxiliary chiral
multiplets $\Phi$ and $\Phi'$ in the $U(N)$ and $U(M)$ 
adjoint representations, respectively. Moreover, 
we introduce two hypermultiplets whose chiral superfields are 
in the bifundamental representations $(\mathbf{N}, \mathbf{M})$
and $(\mathbf{\overline{N}}, \mathbf{\overline{M}})$. 
To be concrete, we denote the bifundamental chiral superfields 
as $(Z_i)_{a \alpha}$ and $(W_i)_{\bar{\alpha} \bar{a}}$ with
$i=1,2$, where $a$ ($\bar{a}$) and $\alpha$ ($\bar{\alpha}$) are 
(anti-)fundamental indices for $U(N)$ and $U(M)$, respectively.
We also choose the Chern-Simons levels of the two gauge groups to be
equal but opposite in sign. The superpotential in our theories is
\begin{eqnarray}
W &=& {k\over {8\pi}} \left({\rm tr} \Phi^{\prime 2} 
- {\rm Tr}\Phi^2\right)
+{\rm tr} \left( W_i \Phi Z_i \right) 
+ {\rm Tr} \left( W_i^T \Phi' Z_i^T\right)~,~\,
\end{eqnarray}
where ${\rm Tr}$ and ${\rm tr}$ are the traces for
$U(N)$ and $U(M)$ gauge groups, respectively, and
the upper index $T$ means transpose.
Because there are no kinetic terms for the auxiliary fields 
$\Phi$ and $\Phi'$, we can integrate them out and obtain the
superpotential 
\begin{eqnarray}
W &=& {{2\pi}\over k} \left( {\rm Tr}(Z_i W_i Z_j W_j)
-{\rm tr}(Z_i^T W_i^T Z_j^T W_j^T)\right)
 \nonumber\\
&=& {{2\pi}\over k}  {\rm Tr} \left(Z_i W_i Z_j W_j
-Z_i W_j Z_j W_i \right)~.~\,
\end{eqnarray}
So, we can rewrite it as follows
\begin{eqnarray}
W &=&  {{2\pi}\over k} \epsilon^{\rho \sigma} 
\epsilon^{\dot{\rho}\dot{\sigma}}
 {\rm Tr} \left(Z_{\rho} W_{\dot{\rho}} Z_{\sigma} 
W_{\dot{\sigma}} \right)~.~\,
\end{eqnarray}
Thus, there are explicit $SU(2)\times SU(2)$ symmetry acting
respectively on the $Z_i$ and $W_i$. 
In addition, we have $SU(2)_R$ symmetry under which the bosonic
components $z_i$ and $w^*_i$ of $Z_i$ and $W^*_i$ 
transform as a doublet.
Because the $SU(2)_R$ symmetry does not commute with the
above $SU(2)\times SU(2)$ symmetry, we can have the 
$SU(4)$ global symmetry under which the four bosonic fields
$(z_1, z_2, w_1^*, w_2^*)$ transform in the fundamental 
${\mathbf{4}}$ representation. The supercharges can not be 
singlets under this $SU(4)$ because $SU(2)_R$ is R-symmetry. 
Note that the generic three-dimensional superconformal theories 
have $SO(N)$ R-symmetry with the supercharge in the
fundamental representation, we should have at least
${\cal N}=6$ superconformal symmetry.

If $N$ is equal to $M$, similar to the above discussions, 
we can construct the three-dimensional ${\cal N}=8$ 
superconformal $SU(N)\times SU(N)$ theories as well. In this case,
we can have another global $U(1)_b$ baryon number symmetry
under which $Z_i$ and $W_i$ have charges $+1$ and $-1$, 
respectively. And we would like to emphasize that 
our $SU(2)\times SU(2)$ theory has enhanced $SO(8)$ R-symmetry
and then is the same as the BLG theory.

Furthermore, our $U(N)\times U(M)$ theories are related to the ABJM 
$U(N)\times U(M)$ theories by the
following transformation on the generators ${\widehat T}^a$ 
of the second gauge group $U(M)$
\begin{eqnarray}
{\widehat T}^a \longrightarrow - {\widehat T}^{aT}~.~\,
\end{eqnarray}

\subsection{ Superconformal $U(1) \times U(1)$ Gauge Theories 
with $SU(2n)$ Global Symmetry}

Similar to the above subsection, we consider the 
${\cal N}=3$ supersymmetric $U(1)\times U(1)$ gauge theories 
and begin from the field content of an ${\cal N}=4$ supersymmetric 
theory.  The auxiliary chiral multiplets 
for the first and second $U(1)$ are $\Phi$ and $\Phi'$ 
respectively. For the matter fields, we consider
two scenarios

(I) We introduce $n$ hypermultiplets whose chiral superfields
have $U(1)\times U(1)$ charges $(\mathbf{+1}, \mathbf{-1})$
and $(\mathbf{-1}, \mathbf{+1})$. 
To be concrete, we denote the bifundamental chiral superfields 
as $(Z_i)_{a \bar{\alpha}}$ and $(W_i)_{\alpha \bar{a}}$ with
$i=1,2, ..., n$, where $a$ ($\bar{a}$) and $\alpha$ ($\bar{\alpha}$) are 
the charge $\mathbf{+1}$ ($\mathbf{-1}$) 
indices for the first and second $U(1)$s, 
respectively. We also choose the Chern-Simons levels of the two gauge 
groups to be equal but opposite in sign. The superpotential  is
\begin{eqnarray}
W &=& {k\over {8\pi}} \left( \Phi^{\prime 2} - \Phi^2\right)
+ W_i \Phi Z_i -  W_i \Phi' Z_i~.~\,
\end{eqnarray}
In general, we only need to require that the Chern-Simons level 
of $U(1)$s be a rational number, {\it i.e.}, $k=p/q$ where $p$ 
and $q$ are relatively coprime.

Integrating out $\Phi$ and $\Phi'$, we obtain
\begin{eqnarray}
W &=& {{2\pi}\over k} \left( Z_i W_i Z_j W_j - Z_i W_i Z_j W_j\right)\equiv 0~.~\,
\end{eqnarray}
Interestingly, the superpotential vanishes. Therefore, we have 
$SU(n) \times SU(n)$ global symmetry acting on $Z_i$ and $W_i$,
respectively. Because $SU(2)_R$ symmetry does not commute with
this $SU(n)\times SU(n)$ symmetry, we should have an $SU(2n)$
global symmetry under which the bosonic fields 
$(z_1, z_2, ..., z_n, w_1^*, w_2^*, ..., w_n^*)$ transform in
the ${\mathbf{2n}}$ fundamental representation of $SU(2n)$.
Note that the supercharges can not be 
singlets under this $SU(4)$ because its subgroup $SU(2)_R$ is 
R-symmetry, we conjecture that
for $n=2$,  the R-symmetry is $SU(4)$ or $SO(6)$;
for $n=3$,  the R-symmetry is $SO(7)$ 
($SU(6) \supset USp(6)\sim SO(7)$);
and for $ 4 \leq n \leq 8$, the R-symmetry is  
$SO(2n)$ or $SO(2n+1)$.
In Section IV, we will derive the $U(1)\times U(1)$ gauge theory 
with $n=4$ from the BLG theory, and then it has
at least $SO(8)$ R-symmetry.

(II) We can introduce $n$ hypermultiplets whose chiral superfields
have $U(1)\times U(1)$ charges $(\mathbf{+1}, \mathbf{+1})$
and $(\mathbf{-1}, \mathbf{-1})$. We denote the bifundamental 
chiral superfields as $(Z'_i)_{a \alpha}$ and 
$(W'_i)_{ \bar{\alpha} \bar{a}}$.
And the Chern-Simons levels of the two gauge 
groups are chosen to be equal but opposite in sign. 
The superpotential  is
\begin{eqnarray}
W &=& {k\over {8\pi}} \left( \Phi^{\prime 2} - \Phi^2\right)
+ W'_i \Phi Z'_i +  W'_i \Phi' Z'_i~.~\,
\end{eqnarray}

Integrating out $\Phi$ and $\Phi'$, we obtain
\begin{eqnarray}
W &=& {{2\pi}\over k} \left( Z'_i W'_i Z'_j W'_j 
- Z'_i W'_i Z'_j W'_j \right)\equiv 0~.~\,
\end{eqnarray}
Because all the rest discussions are the same as those in 
the above scenario, we will not repeat them here.

Furthermore, the $U(1)\times U(1)$ gauge theories with 
$n$ pairs of chiral superfields whose 
$U(1)\times U(1)$ charges are $(\mathbf{\pm 1}, \mathbf{\mp 1})$
are related to  the $U(1)\times U(1)$ gauge theories  with 
$n$ pairs of chiral superfields whose 
$U(1)\times U(1)$ charges are $(\mathbf{\pm 1}, \mathbf{\pm 1})$
via the following transformation
on the gauge field ${\widehat A}_{\mu}$ of the second $U(1)$
\begin{eqnarray}
{\widehat A}_{\mu} \longrightarrow - {\widehat A}_{\mu} ~.~\,
\end{eqnarray}

\section{Relations Among the Three-Dimensional ${\cal N} \geq 5 $
Superconformal Gauge Theories}

Let us briefly review the ABJM theories with generic gauge groups 
$U(N)\times U(M)$ or $SU(N)\times SU(N)$~\cite{Aharony:2008ug}.
Following the convention in Ref.~\cite{Benna:2008zy}, we can write
the explicit Lagrangian  as follows
\begin{eqnarray}
 \label{c}
  {\mathcal L} &=& 
    2 K \epsilon^{\mu\nu\lambda} {\rm Tr} \left(
        A_\mu \partial_\nu A_\lambda 
+ \frac {2i}{3} A_\mu A_\nu A_\lambda
        - \widehat{A}_\mu \partial_\nu \widehat{A}_\lambda 
- \frac {2i}{3} \widehat{A}_\mu \widehat{A}_\nu \widehat{A}_\lambda \right)
     - {\rm Tr} \left(({\mathcal D}_\mu Z)^\dagger {\mathcal D}^\mu Z
\right.\nonumber\\&& \left.
    +  ({\mathcal D}_\mu W)^\dagger {\mathcal D}^\mu W
    + i  \zeta^\dagger \gamma^\mu{\mathcal D}_\mu \zeta
    + i  \omega^\dagger \gamma^\mu{\mathcal D}_\mu \omega \right)
- V_{\mathrm{Bos}} - V_{\mathrm{Ferm}}~,~\,
\end{eqnarray}
 where 
\begin{eqnarray}
K~=~{k\over {8 \pi} }~,~\,
\end{eqnarray}
\begin{eqnarray}
  Z^1  = X^1 + i X^5 ~,~~ ~Z^2  = X^2 + i X^6 ~,~~ 
  ~W_1  = X^{3\dagger} - i X^{7\dagger}  ~,~~ 
  ~W_2  = X^{4\dagger} - i X^{8\dagger}  ~,~\,
\end{eqnarray}
where $X^i$ belongs to the bifundamental $(\mathbf{N}, \mathbf{\overline{M}})$
representation  of $U(N)\times U(M)$, or the bifundamental 
$(\mathbf{N}, \mathbf{\overline{N}})$ representation of 
$SU(N)\times SU(N)$. Also, 
$(Z^i, \zeta^i)$ and $(W_i, \omega_i)$
form chiral superfields. For simplicity, we will not distinguish the trace
between the $U(N)$ and $U(M)$. To write the potential $V_{\mathrm{Bos}}$ and 
$V_{\mathrm{Ferm}}$ that are invariant explicitly under $SU(4)$ R-symmetry,
we define
\begin{eqnarray}
Y^{A}~=~\{Z^A,~ W^{\dagger A}\}~,~~~ Y_A^{\dagger}~=~ \{Z_A^{\dagger}, ~W_A\}~,~\,
\end{eqnarray}
\begin{eqnarray}
\psi_A~=~\{ \epsilon_{AB} \zeta^B e^{-i \pi/4}, ~-\epsilon_{AB} \omega^{\dagger B} 
e^{i \pi/4}\}~,~~~
\psi^{A\dagger}~=~\{ -\epsilon^{AB} \zeta^{\dagger}_B e^{i \pi/4}, 
~\epsilon^{AB} \omega_{ B} 
e^{-i \pi/4}\}~.~\,
\end{eqnarray}
And then, we have 
\begin{eqnarray}
V_{\mathrm{Bos}}&=& -{1\over {48K^2}} {\rm Tr}\left(Y^A Y_A^{\dagger} Y^B Y_B^{\dagger}
Y^C Y_C^{\dagger}+ Y_A^{\dagger} Y^A Y_B^{\dagger} Y^B Y_C^{\dagger} Y^C 
\right.\nonumber\\&& \left.
+ 4 Y^A Y_B^{\dagger} Y^C Y_A^{\dagger} Y^B Y_C^{\dagger}
- 6 Y^A Y_B^{\dagger} Y^B Y_A^{\dagger} Y^C Y_C^{\dagger}\right)~,~\,
\end{eqnarray}
\begin{eqnarray}
V_{\mathrm{Ferm}} &=& {i\over {4K}} {\rm Tr} \left(  Y_A^{\dagger} Y^A
\psi^{B\dagger} \psi_B -  Y^A Y_A^{\dagger} \psi_B \psi^{B\dagger} 
+ 2 Y^A Y_B^{\dagger} \psi_A \psi^{B\dagger} 
- 2  Y_A^{\dagger} Y^B \psi^{A\dagger} \psi_B
\right.\nonumber\\&& \left.
- \epsilon^{ABCD}  Y_A^{\dagger} \psi_B Y_C^{\dagger} \psi_D
+ \epsilon_{ABCD}  Y^A \psi^{B\dagger} Y^C \psi^{D\dagger} \right)~.~\,
\end{eqnarray}

For convention, we choose 
\begin{eqnarray}
{\rm Tr}(T^a T^b)~=~{1\over 2} \delta_{ab}~,~~~[T^a, ~T^b]~=~if_{abc} T^c~,~\,
\end{eqnarray}
where $T^{a, b, c}$ are the generators of the corresponding gauge group.

The Lagrangians for our ${\cal N}=6$ superconformal $U(N)\times U(M)$ 
and $SU(N)\times SU(N)$ 
theories are similar to the above ABJM theories except that  $X^i$ 
belongs to the bifundamental $(\mathbf{N}, \mathbf{{M}})$
representation  for $U(N)\times U(M)$ or the bifundamental 
$(\mathbf{N}, \mathbf{{N}})$ representation  for
$SU(N)\times SU(N)$.
Moreover, we obtain the Lagrangians for the ${\cal N}=5$ superconformal 
$O(N)\times USp(2M)$ gauge theories from the
above  ABJM theories by changing the gauge groups to
 $O(N)\times USp(2M)$ and requiring the reality conditions
\begin{eqnarray}
(W^{A\dagger})_{a {\bar \alpha}} &=& 
\delta_{a{\bar b}} {\omega}_{{\bar \alpha} \beta} 
\epsilon^{AB} (Z^{B*})_{{\bar b} \beta}~,~\,
\end{eqnarray}
where $\omega_{{\bar \alpha} \beta}$ is the anti-symmetric invariant 
tensor of $USp(2M)$.

In the ABJM and our theories, the $U(N)\times U(N)$ theories are
related to the $SU(N)\times SU(N)$ theories. Roughly speaking,
 the ABJM $U(N)\times U(N)$ theories are the direct 
sum of the ABJM $SU(N)\times SU(N)$ theories and our
$U(1)\times U(1)$ theories that have $n=2N^2$ pairs of bifundamental 
chiral superfields with  $U(1)\times U(1)$ charges 
$({\mathbf{\pm1}}, {\mathbf{\mp1}})$. And our
 $U(N)\times U(N)$ theories are the direct
sum of our $SU(N)\times SU(N)$ theories and our
$U(1)\times U(1)$ theories that have $n=2N^2$ pairs of bifundamental 
chiral superfields with  $U(1)\times U(1)$ charges 
$({\mathbf{\pm1}}, {\mathbf{\pm1}})$. Exactly speaking, these statements 
are not completely correct since the matter fields are the same 
matter fields for both
$SU(N)\times SU(N)$ and $U(1)\times U(1)$ theories.
Thus, in the following discussions, we will concentrate on
the $U(N)\times U(M)$ theories proposed by ABJM and us for simplicity.

To study the relations among the ${\cal N} \geq 5$ superconformal 
Chern-Simons gauge theories,
similar to the Wilson line gauge symmetry breaking, we introduce 
a discrete symmetry to the ABJM $U(N)\times U(M)$ theories, 
our $U(N)\times U(M)$ theories, and the ${\cal N}=5$ superconformal
$O(N)\times USp(2M)$ Chern-Simons gauge theories.
Requiring that the theories be invariant under 
the discrete symmetry, we obtain the new three-dimensional superconformal 
Chern-Simons gauge theories. For simplicity, we shall only consider the 
$Z_2$ symmetry in this paper.

\subsection{Relations Among the ${\cal N}=6$ Superconformal 
$U(N)\times U(M)$ Theories}

We consider the ABJM theories and our theories with $U(N)\times U(M)$ 
group. And we introduce the following $Z_2$ transformations that act on the
gauge fields $A_{\mu}$ and $\widehat{A}_\mu$
\begin{eqnarray}
A_{\mu} \longrightarrow \Omega A_{\mu} \Omega^{\dagger}~,~~~
\widehat{A}_\mu \longrightarrow \widehat{\Omega} 
\widehat{A}_{\mu} \widehat{\Omega}^{\dagger}~,~\,
\end{eqnarray}
where $\Omega^2 =1$, and $\widehat{\Omega}^2=1$.
For the ABJM theories, the matter fields
$(Z^i, \zeta^i)$ and $(W_i, \omega_i)$ transform as
\begin{eqnarray}
(Z^i, ~\zeta^i)  \longrightarrow (  \Omega Z^i \widehat{\Omega}^{\dagger},~
\Omega \zeta^i \widehat{\Omega}^{\dagger})~,~~~
(W_i, ~\omega_i)  \longrightarrow ( \widehat{\Omega} W_i  \Omega^{\dagger},~
 \widehat{\Omega} \omega_i \Omega^{\dagger})~.~\,
\end{eqnarray}
And for our $U(N)\times U(M)$ theories, the matter fields 
transform as
\begin{eqnarray}
(Z^i, ~\zeta^i)  \longrightarrow (  \Omega Z^i \widehat{\Omega}^T,~
\Omega \zeta^i \widehat{\Omega}^T)~,~~~
(W_i, ~\omega_i)  \longrightarrow ( \widehat{\Omega}^{T} W_i  \Omega,~
 \widehat{\Omega}^{T} \omega_i \Omega)~.~\,
\end{eqnarray}

To derive the new theories, we choose the
folowing representations for $\Omega$ and $\widehat{\Omega}$ 
\begin{eqnarray}
\Omega~=~\left(
\begin{array}{cc}
 I_{N1\times N1}  &   0\\
0 & -I_{N2\times N2}\\
\end{array}
\right)~,~~~
\widehat{\Omega}~=~\left(
\begin{array}{cc}
 I_{M1\times M1}  &   0\\
0 & -I_{M2\times M2}\\
\end{array}
\right)~,~~~ \,
\end{eqnarray}
where $I_{n\times n}$ is the $n$ by $n$ indentity matrix.
In addition, $N1 > 0$, $N2 \geq 0$, $M1 > 0$, $M2 \geq 0$,
$N1+N2=N$, and $M1+M2=M$.
Therefore, the three-dimensional ${\cal N}=6$ superconformal
Chern-Simons gauge theories with group $U(N)\times U(M)$
 are broken down to two decoupled three-dimensional ${\cal N}=6$ 
superconformal Chern-Simons gauge theories with groups
$U(N1)\times U(M1)$ and $U(N2)\times U(M2)$. 
Also, if $N2=0$ (or $M2=0$), we will have pure Chern-Simons
gauge theories with group $U(M2)$ (or $U(N2)$) which can have any
desired amount of supersymmetry~\cite{Schwarz:2004yj}. 
In short, we have showed that in the ABJM theories and our theories,
the ${\cal N}=6$ superconformal $U(N')\times U(M')$ 
Chern-Simons gauge theories can be obtained from  ${\cal N}=6$ superconformal 
$U(N)\times U(N)$ Chern-Simons gauge theories, and  the ${\cal N}=6$ 
superconformal $U(N')\times U(N')$ Chern-Simons gauge theories can be obtained 
from the ${\cal N}=6$ superconformal $U(N)\times U(M)$ 
Chern-Simons gauge theories, 
where $N' \leq N$,  $M' \leq N$, and $N' \leq M$.

\subsection{Relations between the ${\cal N}=6$ Superconformal
$U(N)\times U(M)$ Chern-Simons Gauge Theories and the ${\cal N}=5$ 
Superconformal $O(N)\times USp(2M)$ Chern-Simons Gauge Theories}

First, we would like to derive the ${\cal N}=5$ superconformal  
$O(N)\times USp(2M)$  Chern-Simons gauge theories from the
${\cal N}=6$ superconformal $U(N)\times U(2M)$ gauge theories
proposed by ABJM and us.
We introduce the following $Z_2$ transformations that act on the
gauge fields $A_{\mu}$ and $\widehat{A}_\mu$
\begin{eqnarray}
A_{\mu} \longrightarrow - A_{\mu}^T~,~~~
\widehat{A}_\mu \longrightarrow J \widehat{A}_{\mu} J~,~\,
\label{OSp-I}
\end{eqnarray}
where 
\begin{eqnarray}
J~=~\left(
\begin{array}{cc}
 0 & I_{M\times M}  \\
 -I_{M\times M} & 0\\
\end{array}
\right)~.~~~ \,
\end{eqnarray}

For the ABJM theories, the matter fields $Y^A$ transform as
\begin{eqnarray}
(Y^A)_{a {\bar \alpha}} \longrightarrow {\widetilde J}^{AB} 
\delta_{a {\bar b}} (Y^{B*})_{{\bar b} {\beta}} 
J_{{\beta}  {\bar \alpha}} ~,~\,
\label{OSp-II}
\end{eqnarray}
and for our theories, the matter fields transform as
\begin{eqnarray}
(Y^A)_{a { \alpha}} \longrightarrow {\widetilde J}^{AB} 
\delta_{a {\bar b}}
(Y^{B*})_{{\bar b} {\bar \beta}} 
J_{{\bar \beta}  {\alpha}} ~,~\,
\label{OSp-III}
\end{eqnarray}
where $a$ and $b$ are $U(N)$ indices, and $\alpha$ and $\beta$ are
$U(2M)$ indices, and 
\begin{eqnarray}
\widetilde J~=~\left(
\begin{array}{cccc}
 0 & 0 & 0  &  1\\
 0 & 0 & -1  &  0\\
 0 & 1 & 0  &  0\\
 -1 & 0 & 0  &  0\\
\end{array}
\right)~.~ \,
\end{eqnarray}
Also, the transformations for $\psi_A$ are similar to those for
$Y^A$. 

Similar to the D-branes on the orientifold planes, we break the
$U(N)\times U(2M)$ gauge symmetry down to the $O(N)\times USp(2M)$
gauge symmetry. In addition, the $SU(4)$ R-symmetry is broken down to 
the $USp(4)$ due to $\widetilde J$ projections. And we only have
a single bifundamental hypermultiplet, or equivalently a pair of 
bifundamental chiral superfields. Therefore,
requiring that the new theory be invariant under the transformations
in Eqs. (\ref{OSp-I}) and  (\ref{OSp-II}) or (\ref{OSp-III}), we obtain the
three-dimensional  ${\cal N}=5$ superconformal $O(N)\times USp(2M)$ 
Chern-Simons gauge theories since $USp(4)\simeq SO(5)$.

Note that $SO(2)\simeq U(1)$, the ${\cal N}=5$ superconformal symmetry 
in the $O(2)\times USp(2M)$ Chern-Simons gauge theories is enhanced to
the ${\cal N}=6$ superconformal symmetry. The point is that
under the gauge symmetry $U(1)\times USp(2M)$, we will have
two bifundamental hypermultiplets, or equivalently two pairs of 
bifundamental chiral superfields. And then, the R-symmetry is
indeed $SU(4)$. Thus, we can derive the ${\cal N}=6$ superconformal
$U(1)\times USp(2M)$ Chern-Simons gauge theories from the ABJM
theories and our theories with gauge group $U(2)\times U(2M)$. 
Although $SO(3)\simeq SU(2)$ and $SO(6)\simeq SU(4)$,
we can show that there is no enhanced $SO(6)$ superconformal symmetry
in the ${\cal N}=5$ superconformal
$O(3)\times USp(2M)$ and $O(6)\times USp(2M)$ gauge theories.

For ${\cal N}=5$ superconformal $G_2\times USp(2)$ Chern-Simons
gauge theory, it seems to us that we might not have the  multiple 
M2-branes'  interpretation. However, we indeed might obtain
such theory from the ${\cal N}=5$ superconformal $O(N)\times USp(2M)$ 
Chern-Simons gauge theories. Note that
$G_2$ is a special maximal subgroup of $SO(7)$, it seems to us that 
the ${\cal N}=5$ superconformal $G_2\times USp(2)$ Chern-Simons 
gauge theory might be derived from the  ${\cal N}=5$ superconformal 
$O(7) \times USp(2)$ Chern-Simons gauge theory.

In short, from the ABJM $U(N)\times U(M)$ theories and  
our $U(N)\times U(M)$ theories,
we can derive the ${\cal N}=5$ superconformal $O(N)\times USp(2M)$ 
Chern-Simons gauge theories and the  ${\cal N}=6$ superconformal 
$U(1)\times USp(2N)$
Chern-Simons gauge theories, and might derive the 
${\cal N}=5$ superconformal 
 $G_2\times USp(2)$ Chern-Simons gauge theory.

Next, we will derive the  ABJM $U(N)\times U(M)$ theories and 
our $U(N)\times U(M)$ theories
from the  ${\cal N}=5$ superconformal $O(2N)\times USp(2M)$ 
Chern-Simons gauge theories. We denote the gauge fields of
$O(2N)$ and $USp(2M)$ as $A_{\mu}$ and ${\widehat A}_{\mu}$, and
the  bifundamental hypermultiplet in terms of two
bifundamental chiral superfields $\Sigma_{a \alpha}$ and 
$\widehat{\Sigma}_{\alpha a}$ where $a$ and $\alpha$ are indices
for $O(2N)$ and $USp(2M)$, respectively. Following the Ref.~\cite{Cahn},
we choose the following generators for $U(N)$ 
\begin{eqnarray}
(T_{ab})_{ij} &=& (e_{a,b})_{ij}\equiv \delta_{ai} \delta_{bj}~,~\,
\end{eqnarray}
where $a$ and $b$ are from 1 to $N$.
In addition, we choose the generators for $SO(2N)$ as follows
\begin{eqnarray}
&& T^1_{ab} ~\equiv~ e_{a,b}-e_{a+N, b+N} ~,~
\nonumber\\&& 
 T^2_{ab} ~\equiv~ e_{a,b+N}-e_{b, a+N} ~,~ ~~~~~~{\rm where}~~a < b~,~
\nonumber\\&& 
 T^3_{ab} ~\equiv~ e_{a+N,b}-e_{b+N, a} ~,~ ~~~~~~{\rm where}~~a < b~,~\,
\end{eqnarray}
where $a$ and $b$ are from 1 to $N$ with possible
extra conditions on $a$ given above.
And we choose the following generators for $USp(2M)$ 
\begin{eqnarray}
&& T^1_{ab} ~\equiv~ e_{a,b}-e_{a+M, b+M} ~,~
\nonumber\\&& 
 T^2_{ab} ~\equiv~ e_{a,b+M}+e_{b, a+M} ~,~ 
\nonumber\\&& 
 T^3_{ab} ~\equiv~ e_{a+M,b}+e_{b+M, a} ~,~ \,
\end{eqnarray}
where $a$ and $b$ are from 1 to $M$.

The fundamental representation of $SO(2N)$ can be decomposed as
the representations of $SU(N)\times U(1)$ as follows
\begin{eqnarray}
{\mathbf{2N}} \longrightarrow \left({\mathbf{N}}, +1\right) \oplus
\left({\mathbf{\overline{N}}}, -1\right)~,~
\end{eqnarray}
where we have normalized the $U(1)$ properly.
Also, the fundamental representation of $USp(2M)$ can be decomposed as
the representations of $SU(M)\times U(1)$ as follows
\begin{eqnarray}
{\mathbf{2M}} \longrightarrow \left({\mathbf{M}}, +1\right) \oplus
\left({\mathbf{\overline{M}}}, -1\right)~.~
\end{eqnarray}

Simiar to the above discussions,
we introduce the following $Z_2$ transformations that act on the 
$O(2N)$ and $USp(2M)$ gauge fields $A_{\mu}$ and $\widehat{A}_\mu$, and 
the matter fields $\Sigma$ and ${\widehat \Sigma}$ 
\begin{eqnarray}
A_{\mu} \longrightarrow \Omega A_{\mu} \Omega~,~~~
\widehat{A}_\mu \longrightarrow {\widehat \Omega} 
\widehat{A}_{\mu} {\widehat \Omega}~,~\,
\label{OSp-XI}
\end{eqnarray}
\begin{eqnarray}
\Sigma \longrightarrow \Omega  \Sigma {\widehat \Omega}~,~~~
{\widehat \Sigma} \longrightarrow {\widehat \Omega}
{\widehat \Sigma} \Omega~.~\,
\label{OSp-XII}
\end{eqnarray}

To derive the ABJM theories with $U(N)\times U(M)$ gauge group,
we choose
\begin{eqnarray}
\Omega~=~\left(
\begin{array}{cc}
 I_{N\times N}  &   0\\
0 & -I_{N\times N}\\
\end{array}
\right)~,~~~
\widehat{\Omega}~=~\left(
\begin{array}{cc}
- I_{M\times M}  &   0 \\
0 & I_{M\times M}\\
\end{array}
\right)~.~~~ \,
\end{eqnarray}
So, the $O(2N)\times USp(2M)$ gauge symmetry is broken down
to the $U(N)\times U(M)$ symmetry. Also, we obtain a pair
of chiral superfields in the $U(N)\times U(M)$ bifundamental 
representations $(\mathbf{N}, \mathbf{\overline{M}})$ and 
$(\mathbf{\overline{N}}, \mathbf{M})$ from $\Sigma$, and another
pair of chiral superfields in the same representations
from $\widehat{\Sigma}$. Therefore, we can have the enhanced 
$SU(4)$ R-symmetry and indeed derive the ${\cal N}=6$ superconformal 
ABJM theories with gauge group $U(N)\times U(M)$.

Furthermore, to derive our theories with $U(N)\times U(M)$ 
gauge group, we choose
\begin{eqnarray}
\Omega~=~\left(
\begin{array}{cc}
 I_{N\times N}  &   0\\
0 & -I_{N\times N}\\
\end{array}
\right)~,~~~
\widehat{\Omega}~=~\left(
\begin{array}{cc}
 I_{M\times M}  &   0 \\
0 & - I_{M\times M}\\
\end{array}
\right)~.~~~ \,
\end{eqnarray}
Thus, the $O(2N)\times USp(2M)$ gauge symmetry is broken down
to the $U(N)\times U(M)$ symmetry. Moreover, we obtain a pair
of chiral superfields in the $U(N)\times U(M)$ bifundamental 
representations $(\mathbf{N}, \mathbf{M})$ and 
$(\mathbf{\mathbf{\overline{N}}}, \mathbf{\overline{M}})$ from $\Sigma$, 
and another pair of chiral superfields in the same representations
from $\widehat{\Sigma}$. Therefore, we can also have the enhanced 
$SU(4)$ R-symmetry and  derive our ${\cal N}=6$ superconformal theories 
with gauge group $U(N)\times U(M)$.

\section{Three-Dimensional ${\cal N}=8$ Superconformal 
$U(1) \times U(1)$ Theory from the BLG Theory}

We shall briefly review the BLG theory in the 
product gauge group formulation by 
van Raamsdonk~\cite{VanRaamsdonk:2008ft}.
In this formulation, the BLG theory is rewritten
as a superconformal Chern-Simons theory with the
$SU(2)\times SU(2)$ gauge group and bifundamental matters,
which has a manifest global $SO(8)$ R-symmetry.
The Lagrangian is  given by~\cite{VanRaamsdonk:2008ft} 

\begin{eqnarray}
{\cal L} &=& 
 {\rm Tr} \left(\frac{1}{2f} \epsilon^{\mu\nu\lambda}(A_\mu \partial_\nu A_\lambda 
+ \frac{2i}{3} A_\mu A_\nu A_\lambda)
    - \frac{1}{2f} \epsilon^{\mu\nu\lambda}(\widehat{A}_\mu \partial_\nu 
\widehat{A}_\lambda
 + \frac{2i}{3} \widehat{A}_\mu \widehat{A}_\nu \widehat{A}_\lambda)
\right.\nonumber\\&& \left.
- ({\mathcal D}^\mu X^I)^\dagger {\mathcal D}_\mu X^I  
+  i\bar{\Psi}^\dagger \Gamma^\mu {\mathcal D}_\mu \Psi 
-\frac{8f^2}{3}  X^{[I} X^{\dagger J} 
X^{K]} X^{\dagger[K} X^{J} X^{\dagger I]} 
\right.\nonumber\\&& \left.
    -\frac{2if}{3} \bar{\Psi}^\dagger \Gamma_{IJ} 
(X^I X^{J\dagger} \Psi + X^J \Psi^{\dagger} X^I 
+ \Psi X^{I\dagger} X^J) \right)~,~\,
\label{eqn:BL-action}
\end{eqnarray}
where the fermions $\Psi$ are represented by 32-component
Majorana spinors of $SO(1, 10)$ subject to
a chirality condition on the world volume which
leaves 16 real degrees of freedom. And
 the covariant derivative is
\begin{eqnarray}
\label{eqn:cov-derivative}
  {\mathcal D}_\mu X = \partial_\mu X + i A_\mu X - i X \widehat{A}_\mu \; .
\end{eqnarray}
The Chern-Simons level $k$ is related to $f$ as follows
\begin{eqnarray}
  f = \frac{2\pi}{k} \; .
\end{eqnarray}

The bifundamental scalars $X^I$ are related to the original BLG variables $x_a^I$ 
with $SO(4)$ index $a$ through
\begin{eqnarray}
\label{eqn:map-so4-su2su2}
  X^I = {1\over 2} (x_4^I ~I_{2\times 2} + i ~x_i^I ~\sigma^i) ~,~\,
\end{eqnarray}
where $\sigma^i$ are the Pauli matrices. And there is a similar 
expression for spinor $\Psi$. Also, 
 the scalars need to satisfy the reality condition
\begin{eqnarray}
\label{eqn:reality-condition}
  X^I_{a {\bar \alpha}}  ~= ~\epsilon_{a {\bar b}} ~
\epsilon_{{\bar \alpha} \beta} ~
(X^{I\dagger})^{\beta {\bar b}}  ~.~\,
\end{eqnarray}
To be concrete, we define $X^I$ as following
\begin{eqnarray}
X^I~\equiv~{1\over {\sqrt 2}}\left(
\begin{array}{cc}
 z^I  &   w^I\\
-{\bar w}^I & {\bar z}^I\\
\end{array}
\right)~,~\,
\end{eqnarray}
where
\begin{eqnarray}
z^I~=~ {1\over {\sqrt 2}} (x_4^I + i x_3^I)~,~~~
w^I~=~ {1\over {\sqrt 2}} (x_2^I + i x_1^I)~.~\,
\end{eqnarray}

Moreover,  the $SU(2)\times SU(2) $ gauge transformations are 
\begin{eqnarray}
\label{eqn:gauge-trafo-components}
  A_\mu   \rightarrow U A_\mu U^\dagger - i U \partial_\mu U^\dagger ~,~~~ 
 \widehat{A}_\mu   \rightarrow \widehat{U} \widehat{A}_\mu \widehat{U}^\dagger 
- i \widehat{U} \partial_\mu \widehat{U}^\dagger ~,~~~
 X^I \rightarrow U X^I \widehat{U}^\dagger ~.~\,
\end{eqnarray}

Similar to the last Section, we consider the $Z_2$ discrete symmetry.
Under $Z_2$ symmetry,  the gauge fields and matter fields transform 
as follows
\begin{eqnarray} 
 A_\mu   \rightarrow \Omega A_\mu \Omega^\dagger ~,~~~ 
 \widehat{A}_\mu   \rightarrow \widehat{\Omega} \widehat{A}_\mu 
\widehat{\Omega}^\dagger  ~,~~~
 X^I \rightarrow \Omega X^I \widehat{\Omega}^\dagger ~.~\,
\end{eqnarray}
The transformations for $\Psi$ are similar to these for $X^I$.
And there are three kinds of independent and non-trivial representations 
for $\Omega$ and $\widehat{\Omega}$
\begin{eqnarray}
\Omega~=~\left(
\begin{array}{cc}
 1  &   0\\
0 & 1\\
\end{array}
\right)~,~~~
\widehat{\Omega}~=~\left(
\begin{array}{cc}
 1 &   0\\
0 & -1\\
\end{array}
\right)~,~~~ \,
\label{BLG-Tr-I}
\end{eqnarray}
\begin{eqnarray}
\Omega~=~\left(
\begin{array}{cc}
 1  &   0\\
0 & -1\\
\end{array}
\right)~,~~~
\widehat{\Omega}~=~\left(
\begin{array}{cc}
 1 &   0\\
0 & 1\\
\end{array}
\right)~,~~~ \,
\label{BLG-Tr-II}
\end{eqnarray}
\begin{eqnarray}
\Omega~=~\left(
\begin{array}{cc}
 1  &   0\\
0 & -1\\
\end{array}
\right)~,~~~
\widehat{\Omega}~=~\left(
\begin{array}{cc}
 1 &   0\\
0 & -1\\
\end{array}
\right)~.~~~ \,
\label{BLG-Tr-III}
\end{eqnarray}
For the first case with $\Omega$ and $\widehat{\Omega}$ in 
Eq. (\ref{BLG-Tr-I}), the gauge group $SU(2)\times SU(2)$ is
 broken down to $SU(2)\times U(1)$. In addition,
the components $w^I$ and ${\bar z}^I$ of $X^I$ are projected
out. So the reality condition in Eq. ({\ref{eqn:reality-condition}}) 
can not be satisfied. Note that only the components 
$z^I$ and ${\bar w}^I$ of $X^I$ can not form 
the complete chiral superfields under the remaining
gauge symmetry $SU(2)\times U(1)$. In order to have the complete
complex representation under $SU(2)\times U(1)$,  we redefine 
\begin{eqnarray}
Y^i~\equiv~\left(
\begin{array}{c}
z^i+iz^{i+4} \\
-{\bar w}^i+i{\bar w}^{i+4}\\
\end{array}
\right)~,~\,
\end{eqnarray}
where $i=$1, 2, 3, 4. 
The discussions for the spinor are similar. Thus, we have four 
chiral superfields in the bifundamental representation of $SU(2)\times U(1)$.
And then, we obtain the three-dimensional ${\cal N}=6$ superconformal
$SU(2)\times U(1)$ theory, {\it i.e.}, the ABJM theory and our theory
with group $SU(2)\times U(1)$
since these two theories are identical in this case.
The discussions for the second case with $\Omega$ and $\widehat{\Omega}$ in 
Eq. (\ref{BLG-Tr-II}) are the same as the first case since 
they are related by the parity symmetry.

The most interesting case is the third one with $\Omega$ and 
$\widehat{\Omega}$ in Eq. (\ref{BLG-Tr-III}). We break the
$SU(2)\times SU(2)$ gauge symmetry down to its Cartan subgroup
$U(1)\times U(1)$, and we project out the components 
$w^I$ and ${\bar w}^I$. Thus, we have eight complex scalar fields
$z^I$ with charge $(\mathbf{+1}, \mathbf{-1})$ 
under gauge group $U(1)\times U(1)$.
The discussions for the spinor $\Psi$ are similar, and we get
a complex Dirac fermion $\psi$ with  charge 
$(\mathbf{+1}, \mathbf{-1})$ under 
gauge group $U(1)\times U(1)$. Similar to 
Ref.~\cite{Benna:2008zy}, we can rewrite 
$(z^i, \psi)$ as 8 chiral superfields, which will not be
discussed here. Because we do not break the supersymmetry,
we obtain the  three-dimensional ${\cal N}=8$ superconformal
$U(1)\times U(1)$ gauge theory. And the Lagrangian is
\begin{eqnarray}
{\cal L} &=& 
  \frac{1}{4f} \epsilon^{\mu\nu\lambda}\left(
A^3_\mu \partial_\nu A^3_\lambda 
    - \widehat{A}^3_\mu \partial_\nu \widehat{A}^3_\lambda\right)
- ({\mathcal D}^\mu z^I)^\dagger {\mathcal D}_\mu z^I  
+  i{\psi}^\dagger \Gamma^\mu {\mathcal D}_\mu \psi ~,~\,
\label{eqn:TL-action}
\end{eqnarray}
where the covariant derivative is
\begin{eqnarray}
  {\mathcal D}_\mu z^I = \partial_\mu z^I + {i\over 2} 
(A^3_\mu -  \widehat{A}^3_\mu) z^I ~.~\,
\end{eqnarray}

We define 
\begin{eqnarray}
 a_{\mu}~\equiv~ {1\over 2} \left(A_{\mu}^3 + \widehat{A}^3_\mu\right)~,~~~
\hat{a}_{\mu}~\equiv~ {1\over 2} \left(A_{\mu}^3 - \widehat{A}^3_\mu\right)~.~\,
\end{eqnarray}
And then we rewrite the above Lagrangian in Eq. (\ref{eqn:TL-action})
as follows
\begin{eqnarray}
{\cal L} &=& 
  \frac{1}{2f} \epsilon^{\mu\nu\lambda} \hat{a}_{\mu} f_{\nu \lambda}
- ({\mathcal D}^\mu z^I)^\dagger {\mathcal D}_\mu z^I  
+  i{\psi}^\dagger \Gamma^\mu {\mathcal D}_\mu \psi ~,~\,
\label{eqn:UI-action}
\end{eqnarray}
where 
\begin{eqnarray}
f_{\mu\nu}~=~\partial_{\mu} a_{\nu} -\partial_{\nu} a_{\mu}~,~~~
  {\mathcal D}_\mu z^I ~=~ \partial_\mu z^I + i \hat{a}_{\mu} z^I ~.~\,
\end{eqnarray}
So, only the $\hat{a}_{\mu}$ gauge field couples to the matter fields.
In particular, because the gauge symmetry in our theory is 
$U(1)\times U(1)$, the level $k$ can be a rational number in general, 
{\it i.e.}, $k=p/q$ where $p$ and $q$ are relatively prime.
We would like to point out that this $U(1)\times U(1)$ gauge theory 
derived from the BLG theory is the same as 
 our $U(1)\times U(1)$ theory with 
four pairs of chiral superfields whose 
$U(1)\times U(1)$ charges are $(\mathbf{\pm 1}, \mathbf{\mp 1})$.

\subsection{Moduli Space}

We will study the moduli space of our $U(1)\times U(1)$ theory,
and focus on the gauge fields and scalar fields.
The discussions are similar to those in 
Refs.~\cite{Lambert:2008et, Distler:2008mk}.  
The gauge transformations for $A^3_\mu$
and $\widehat{A}^3_\mu$ are
\begin{eqnarray}
A^3_\mu \longrightarrow A^3_\mu -\partial_{\mu} \theta~,~~~
\widehat{A}^3_\mu \longrightarrow \widehat{A}^3_\mu 
-\partial_{\mu} \hat{\theta} ~,~\,
\end{eqnarray}
so we obtain the gauge transformations for $a_{\mu}$, $\hat{a}_{\mu}$,
and $z^I$ 
\begin{eqnarray}
z^I \longrightarrow e^{i(\theta-\hat{\theta})/2} z^I~,~~~
a_{\mu} \longrightarrow a_{\mu} 
-{1\over 2} (\partial_{\mu}\theta + \partial_{\mu}\hat{\theta})~,~~~
\hat{a}_{\mu}  \longrightarrow  \hat{a}_{\mu}
-{1\over 2} (\partial_{\mu}\theta-\partial_{\mu}\hat{\theta})~.~\,
\end{eqnarray}
We define 
\begin{eqnarray}
\alpha ~\equiv~ {1\over 2} (\theta+\hat{\theta})~,~~~
\hat{\alpha} ~\equiv~ {1\over 2} (\theta-\hat{\theta})~,~\,
\end{eqnarray}
Then we obtain the gauge transformations for $a_{\mu}$, $\hat{a}_{\mu}$,
and $z^I$
\begin{eqnarray}
z^I \longrightarrow e^{i\hat{\alpha}} z^I~,~~~
a_{\mu} \longrightarrow a_{\mu} -\partial_{\mu} \alpha ~,~~~
\hat{a}_{\mu}  \longrightarrow  \hat{a}_{\mu}
-\partial_{\mu} \hat{\alpha}~.~\,
\label{Gauge-Tr}
\end{eqnarray}
We emphasize that both $\alpha$ and $\hat{\alpha}$ have period $2\pi$,
which is consistent with the $2\pi$ period for $z^I$.

Moreover, we introduce a Lagrange multiplier term
\begin{eqnarray}
{\cal L}_{\sigma}~=~ {1\over {4\pi}} \sigma 
\epsilon^{\mu\nu\lambda} \partial_{\mu} f_{\nu \lambda}  ~,~\,
\end{eqnarray}
where $\sigma$ is a new scalar field and has period $2\pi$. 
And then we can treat $f_{\mu\nu}$ as an independent variable.

Therefore, the relevant Lagrangian for the gauge
and scalar fields is
\begin{eqnarray}
{\cal L}~=~ -|\partial_{\mu} z^I + i \hat{a}_{\mu} z^I|^2
+{k\over {4\pi}}\epsilon^{\mu \nu \lambda} \hat{a}_{\mu} f_{\nu \lambda}
+{1\over {4\pi}} \sigma \epsilon^{\mu \nu \lambda} \partial_{\mu} 
f_{\nu \lambda}~.~\,
\end{eqnarray}
From the equation of motion
for $f_{\mu\nu}$, we have
\begin{eqnarray}
\hat{a}_{\mu} ~=~ {1\over k} \partial_{\mu} \sigma~.~\,
\label{EOM-f}
\end{eqnarray}
Using this, we obtain the reduced action
\begin{eqnarray}
{\cal L}~=~ -|\partial_{\mu} z^I + {i\over k} 
(\partial_{\mu} \sigma) z^I|^2 ~.~\,
\end{eqnarray}
From the Eqs. (\ref{Gauge-Tr}) and (\ref{EOM-f}), we obtain
the new gauge transformations for $z^I$ and $\sigma$
\begin{eqnarray}
z^I \longrightarrow e^{i\hat{\alpha}} z^I~,~~~
\sigma \longrightarrow \sigma - k \hat{\alpha}~.~
\end{eqnarray}
Fixing the gauge $\sigma=0$, we still have the residual
gauge transformation
\begin{eqnarray}
\hat{\alpha} (x)~=~ {{2n\pi }\over k}~.~\,
\end{eqnarray}
Thus, the moduli space is characterized by a set of
eight complex numbers $z^I$, and the residual symmetry 
transformations are
\begin{eqnarray}
z^I \longrightarrow e^{i2n\pi/k} z^I~,~~~
\end{eqnarray}
and 
\begin{eqnarray}
z^I \longrightarrow  {\bar z}^I~.~~~
\end{eqnarray}
Therefore, we conclude that for $k=1$, the moduli space is
\begin{eqnarray}
(\mathbb{R}^8 \times \mathbb{R}^8)/Z_2 ~,~\,
\end{eqnarray}
for $k=2$, the moduli space is
\begin{eqnarray}
(\mathbb{R}^8 \times \mathbb{R}^8)/(Z_2 \times Z_2) ~,~\,
\end{eqnarray}
and for $k>2$, the moduli space is
\begin{eqnarray}
(\mathbb{R}^8 \times \mathbb{R}^8)/{\rm Dih}_{k} ~,~\,
\end{eqnarray}
where ${\rm Dih}_{k}$ is the dihedral group.

For $k=p/q$ in general, the discussions of moduli space 
 are similar to the above, so we will not present them here. 

\subsection{Novel Higgs Mechanism}

Similar to the Refs.~\cite{Mukhi:2008ux, Li:2008ya}, 
we can obtain the D-brane action
via the novel Higgs mechanism. For simplicity, we focus
on the gauge fields here.
Because we have the $SU(8)$ global symmetry, we can always 
make the rotation so that
only the scalar field $ z^8  $ develops a 
vacuum expectation value (VEV)
\begin{eqnarray}
 \langle z^8 \rangle ~=~v ~.~\,
\end{eqnarray}
Then $\hat{a}_{\mu}$ becomes massive, and we obtain the 
relevant Lagrangian for gauge fields
\begin{eqnarray}
{\cal L} &=& {k\over {4\pi}} \epsilon^{\mu\nu\lambda} 
\hat{a}_{\mu} f_{\nu \lambda} - v^2 \hat{a}^2_{\mu} ~,~\,
\end{eqnarray}
where the second term comes from the kinetic term of $z^8$.
Integrating out $\hat{a}_{\mu}$, we obtain
\begin{eqnarray}
{\cal L} &=& - {{k^2}\over {32\pi^2v^2}} f^{\mu \nu} f_{\mu \nu}
 ~.~\,
\label{Action-amu}
\end{eqnarray}
Thus, the gauge field $a_{\mu}$ becomes a dynamical
field, and its gauge coupling is
\begin{eqnarray}
g &=& 2 {\sqrt 2} \pi {v\over k} ~.~\,
\end{eqnarray}
So, for very larg $v$ and $k$, we can still 
fix the gauge coupling $g$ as a constant.

Because we have eight chiral superfiels
from $z^I$ and $\psi$, we should have another
decoupled $U(1)$ gauge field in addition to
$a_{\mu}$ if we only have ${\cal N}=8$ superconformal
symmetry. To be concrete, we shall choose
the unitary gauge, and define  $z^8$ as
follows
\begin{eqnarray}
z^8 & \equiv & {1\over {\sqrt 2}} \rho e^{i \phi} ~.~\,
\end{eqnarray}
From the kinetic term of $z^8$, we obtain the
kinetic term for $\rho$
\begin{eqnarray}
 {\cal L} & = & -{1\over 2} |\partial_{\mu} \rho 
+ iB_{\mu} \rho|^2  ~,~\,
\end{eqnarray}
where
\begin{eqnarray}
B_{\mu} &=& \hat{a}_{\mu} + \partial_{\mu} \phi  ~.~\,
\end{eqnarray}
And the Chern-Simons term for the gauge fields becomes
\begin{eqnarray}
{k\over {4\pi}}\epsilon^{\mu \nu \lambda} \hat{a}_{\mu} f_{\nu \lambda}
&=& {k\over {4\pi}} \epsilon^{\mu \nu \lambda}
\left( B_{\mu}  f_{\nu \lambda} - 
\partial_{\mu} \phi  f_{\nu \lambda} \right)~.~\,
\end{eqnarray}
Using the Bianchi identity for $f_{\nu \lambda}$,
the last term in the above Lagrangian vanishes.
Giving the following VEV to $\rho$
\begin{eqnarray}
 \langle \rho \rangle ~=~{\sqrt 2} v ~,~\,
\end{eqnarray}
we obtain the  
relevant Lagrangian for gauge fields
\begin{eqnarray}
{\cal L} &=& {k\over {4\pi}} \epsilon^{\mu\nu\lambda} 
B_{\mu} f_{\nu \lambda} - v^2 B^2_{\mu} ~.~\,
\end{eqnarray}
Integrating out the gauge field $B_{\mu}$, we obtain
the same action for $a_{\mu}$ given in Eq. (\ref{Action-amu}).

Therefore, the physical picture is the following: 
the field $\phi$ is eaten by the gauge field $B_{\mu}$,
and the gauge field $a_{\mu}$ becomes dynamical after
$B_{\mu}$ is integrated out. Equivalently speaking,
 the field $\phi$ is eaten by the gauge field $a_{\mu}$.
In addition, $\rho$ is dual to the decoupled $U(1)$ gauge
field, which explains why we have eight chiral 
superfields from $z^I$ and $\psi$.
In the unitary gauge, similar discussions can be
applied to the novel Higgs mechanism in all the
 ${\cal N} \geq 5$ superconformal Chern-Simons 
gauge theories.

\section{Discussion and Conclusions}

In this paper, we have proposed the three-dimensional 
${\cal N}=6$ superconformal $U(N)\times U(M)$ gauge theories
with two pairs of bifundamental chiral superfields in 
the $(\mathbf{N}, \mathbf{M})$ and  
$(\mathbf{\overline{N}}, \mathbf{\overline{M}})$ representations,
and the ${\cal N}=6$ superconformal $SU(N)\times SU(N)$
gauge theories with two pairs of bifundamental chiral superfields in 
the $(\mathbf{N}, \mathbf{N})$ and  
$(\mathbf{\overline{N}}, \mathbf{\overline{N}})$ representations.
For our $SU(N)\times SU(N)$ theory with $N=2$, we reproduced the BLG theory.
We also proposed the superconformal $U(1)\times U(1)$ gauge 
theories that have $n$ pairs of bifundamental chiral superfields with 
the  $U(1)\times U(1)$ charges $({\mathbf{\pm1}}, {\mathbf{\mp1}})$,
or the $U(1)\times U(1)$ charges $({\mathbf{\pm1}}, {\mathbf{\pm1}})$.
Although these theories have global 
symmetry  $SU(2n)$,  the R-symmetry is $SO(6)$ for $n=2$ and
might be $SO(2n)$ or $SO(2n+1)$ for $3 \le n \leq 8$.

In addition,  we showed that in the ABJM $U(N)\times U(M)$ theories and 
our $U(N)\times U(M)$ theories, the ${\cal N}=6$ superconformal $U(N')\times U(N')$ 
Chern-Simons gauge theories can be obtained from the ${\cal N}=6$ superconformal 
$U(N)\times U(M)$ Chern-Simons gauge theories,  and vice versa.
Moreover, we proved that the ${\cal N}=5$ superconformal 
$O(N) \times USp(2M)$ Chern-Simons gauge theories can be derived from 
the  ABJM $U(N)\times U(2M)$ theories and our $U(N)\times U(2M)$ theories.
Also, both the ABJM $U(N)\times U(M)$ theories and 
our $U(N)\times U(M)$ theories can be derived from the 
${\cal N}=5$ superconformal $O(2N) \times USp(2M)$ Chern-Simons gauge 
theories as well. Moreover, we explained that the $SO(5)$ R-symmetry in the 
  ${\cal N}=5$ superconformal $O(2)\times USp(2N)$ gauge theories is
enhanced to $SO(6)$, and then we obtained the ${\cal N}=6$ superconformal 
$U(1)\times USp(2N)$ gauge theories. Because $G_2$ is a special maximal 
subgroup of $SO(7)$, it seems to us that the ${\cal N}=5$ superconformal 
$G_2\times USp(2)$ Chern-Simons gauge theory might be obtained from 
the  ${\cal N}=5$ superconformal $O(7) \times USp(2)$ Chern-Simons gauge
theory.

Furthermore, we derived the three-dimensional  ${\cal N}=8$ superconformal 
$U(1) \times U(1)$ gauge theory from the BLG theory, which can be
considered as our  superconformal $U(1)\times U(1)$ gauge theory
with four pairs of chiral superfields whose
 $U(1)\times U(1)$ charges are $({\mathbf{\pm1}}, {\mathbf{\mp1}})$.
Also, we have studied
the  moduli space in details. Considering the
novel Higgs mechanism in the unitary gauge, we 
  suggested that this theory may describe 
a D2-brane and a decoupled D2-brane.

\begin{acknowledgments}

 This research was supported in part  by the
Cambridge-Mitchell Collaboration in Theoretical Cosmology.

\end{acknowledgments}

\renewcommand{\Large}{\large}

\end{document}